# Astro2020 Science White Paper

# Technosignatures in Transit

**Principal Author: Jason T. Wright**
astrowright@gmail.com
525 Davey Laboratory
Department of Astronomy & Astrophysics and Center for Habitable Worlds
Penn State University
University Park, PA 16802
(814) 863-8470

**Co-author**: **David Kipping**, Columbia University

**Endorsers / co-signers: Daniel Angerhausen,** Center for Space and Habitability, Bern University, Switzerland; Blue Marble Space Institute of Sciences, Seattle, USA; **Tabetha Boyajian** Louisiana State University; **Douglas A. Caldwell**, SETI Institute

**Abstract:** *Kepler*, *K2, TESS*, and similar time-domain photometric projects, while designed with exoplanet detection in mind, are also well-suited projects for searches for large artificial structures orbiting other stars in the Galaxy. An effort to examine these data sets with an eye towards non-spherical or otherwise anomalous transit events, and a robust follow-up program to understand the stars and occulters that generate them, would enable the first robust upper limits on such megastructures in terms of their sizes, occurrence rates, and orbital properties. Such work also has the ancillary benefit of improving our understanding of stellar photometric variability and orbital and physical parameter estimation of exoplanets from photometric time series, and may lead to the identification of new, unexpected classes of stellar variables and exoplanets. Ultimately, **searching for the most unusual and anomalous signatures benefits not only the search for technologies, but also the entire astronomical community by uncovering new mysteries to advance our understanding of the Universe**.

# The Rationale for Searching for Orbital Technosignatures

Since "technology" might be defined as the deliberately engineered applications of energy, some of the most general technosignatures are those related to energy use. It is reasonable to suppose that such technology might be found in space, for instance in the form of satellites and space probes like our own.

Since technological life elsewhere in the Galaxy is unlikely to be as young as our own species' technology, there has been time for its technology to have very large scope, perhaps involving extraordinarily large numbers of small objects or even planet-sized objects and/or collections of objects ("megastructures"). Expansion on this scale need not have a single purpose or deliberate design behind it: humanity's megalopolises are in some sense large, heterogeneous structures that grew more-or-less organically over centuries or millennia. Similarly, a driver as simple as collecting more energy for some purpose might be sufficient, over the course of millions of years, to create a swarm of solar-panel-bearing objects around a star.

Even if stellar energy collection is not the purpose of such structures, fundamental principles of engineering require that they manage the thermal consequences of their energy use, as required by the laws of thermodynamics, which in space (where only radiative cooling is efficient) means having large radiative surfaces, proportional in area to the power consumption of the technology.

Such a swarm (sometimes called a "Dyson sphere", Dyson 1960) would be detectable as a significant infrared excess, and this is the subject of some searchers for technosignatures in the infrared (see the companion white paper by J. Wright). But a second way to detect such structures is by the starlight they intercept, in particular as they transit their host star.

Searching for the transit signatures of orbiting structures satisfies many of the desiderata of a good SETI search: the signature is at least as long-lived as the underlying technology, large collecting/radiating areas are an inevitable consequence of all energetic technology use, and it is detectable not only with current astronomical instrumentation, but indeed much of the data needed to perform the search has been or will be taken in the search for exoplanets.

The primary drawback of the method is that unlike communicative transmissions, transit events (and similar dimmings) of stars are common for many natural reasons. This, however, also means that searches for anomalous dimming events has good synergies with stellar and exoplanetary astronomy, while still providing other forms of SETI with enriched target lists for more unambiguous technosignatures. Indeed it satisfies Freeman Dyson's "First Law of SETI Investigations": "every search for alien civilizations should be planned to give interesting results even when no aliens are discovered".

A search for the truly weird and most anomalous stars in the cosmos would complement other technosignature work by relaxing the need to predict extraterrestrial behavior. Most importantly, a "search for the weird" benefits all of astronomy by uncovering objects which defy our current understanding. Their discovery will drive theoretical research and motivate new observational programs, expanding our understanding to encompass the new phenomena. Establishing that the cause is truly artificial will likely be a long and difficult road, with a rich enough reward to justify the work.

# Towards Robust Upper Limits of Orbiting Technosignatures

Interpreting transit measurements of stars serves searches for technosignatures in two ways: 1) it identifies anomalous infrared sources for further study as both natural or artificial objects of interest (Djorgovski 2000), and 2) it allows for upper limits to be computed—that is, the identification of the regions of parameter space which have been excluded in the search for alien industry (e.g. Wright, Kanodia, & Lubar 2018).

*Kepler* ushered in a new era of stellar and exoplanetary astrophysics by demonstrating the feasibility and utility of precise photometric monitoring of large numbers of stars. The primary purpose was of course to search for exoplanets, but the principle of searching for transiting artificial structures is identical, and indeed *Kepler* inadvertently was a nearly perfect survey instrument for such work.

*Kepler, K2*, and now *TESS* and soon other observatories thus provided the data necessary to compute the first rigorous upper limits on transiting structures, in a manner similar to the exoplanet population demographics it has produced. The primary tasks of such work are:
1. Computing the sensitivity of these surveys as a function of megastructure properties

2. Identifying signals in the light curves consistent with megastructures but not with common, natural phenomena such as intrinsic stellar variability, disks, or exoplanets
3. Following up anomalous stars to determine their nature

The first two of these tasks were explored by Wright et al. (2016), who identified ten classes of transit anomaly that could distinguish them from natural phenomena, as shown in Table 1.

**Table 1**
Ten Anomalies of Transiting Megastructures that Could Distinguish Them from Planets or Stars

| Anomaly | Artificial Mechanism | Natural Confounder |
|---|---|---|
| Ingress and egress shapes | non-disk aspect of the transiting object or star | exomoons, rings, planetary rotation, gravity and limb darkening, evaporation, limb starspots |
| Phase curves | phase-dependent aspect from non-spherical shape | clouds, global circulation, weather, variable insolation |
| Transit bottom shape | time-variable aspect turing transit, e.g., changes in shape or orientation | gravity and limb darkening, stellar pro/oblateness, starspots, exomoons, disks |
| Variable depths | time-variable aspect turing transit, e.g., changes in shape or orientation | evaporation, orbital precession, exomoons |
| Timings/durations | non-gravitational accelerations, co-orbital objects | planet–planet interactions, orbital precession, exomoons |
| Inferred stellar density | non-gravitational accelerations, co-orbital objects | orbital eccentricity, rings, blends, starspots, planet–planet interactions, very massive planets |
| Aperiodicity | Swarms | Very large ring systems, large debris fields, clumpy, warped, or precessing disks |
| Disappearance | Complete obscuration | clumpy, warped, precessing, or circumbinary disks |
| Achromatic transits | Artifacts could be geometric absorbers | clouds, small-scale heights, blends, limb darkening |
| Very low mass | Artifacts could be very thin | large debris field, blends |

Briefly, unlike planets, megastructures orbiting stars might be nonspherical and non-circular, low mass and potentially subject to radiation pressure, opaque, in orbits uncharacteristic of planets, or much larger than planets. All of these properties can be parameterized in terms of light curve properties, allowing us to set both single-target and population upper limits on various classes of megastructure, such as the exo-Clarke belt proposed by Soccas-Novarro (2018). Similar upper limits have been calculated for, e.g., exomoons (e.g. Kipping et al. 2012).

# A Case Study: Boyajian's Star

A good illustration of step 3 at the top of this page is the case of Boyajian's Star. Wright et al. (2016) called out the unusual light curve of Boyajian's Star as a particularly strange and inexplicable example of an anomalous object whose nature need be established as natural before upper limits could be computed. The star might have gone unnoticed for some time without the efforts of the Planet Hunters citizen science team, which flagged the light curve as strange, and the dedicated efforts of Boyajian et al. (2016) who showed that the system was not young and exhibited none of the usual properties of "dipper" stars being occulted by giant planetary ring systems (e.g. Kenworthy & Mamajek 2015) or disks. The star later proved to have many other anomalous photometric properties (e.g. Montet & Simon 2016).

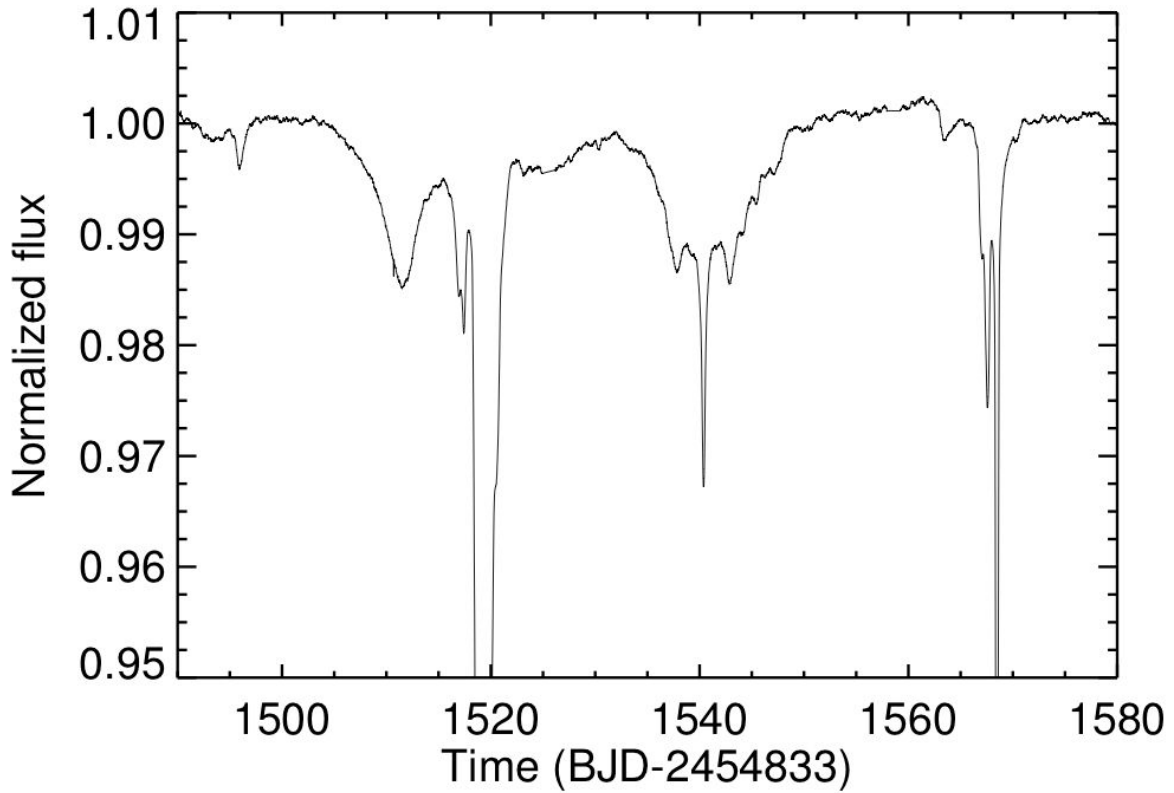

*Figure 1: The photometric light curve of Boyajian's Star from* Kepler, *showing what appear to be a series of occulters of various sizes, shapes, and orbital speeds.*

The ensuing flurry of activity produced many hypotheses (summarized in Wright & Sigurdsson 2016). The subsequent detection of reddening during a later dip observed from the ground confirmed that the occulting material is consistent with dust, not opaque objects (Bodman et al. 2018), thus putting to bed the megastructure hypothesis. But the star continues to elude a single consensus explanation. **Boyajian's star has thus produced a significant amount of theory and observation in both the fields of stellar astrophysics and technosignature search. A search for similar, less obvious objects should prove similarly fruitful on both fronts.**

# A Weird Filter and Inverting Light Curves

Identifying weird stars through human inspection, as was done for Boyajian's Star will become increasingly impractical as our datasets grow in volume. Further, humans are much harder to statistically calibrate making the interpretation of such surveys less rigorous. Automated approaches have started to appear in the literature, such as the use of unsupervised machine learning for identifying anomalous events (Giles & Walkowicz 2018) and phase-dispersion minimization for identifying periodic anomalous events (Technosignature Workshop Participants 2019, p.26). The continued development of such resources, including workshops focussed on collaborative development, could change the shape of technosignature research.

But merely identifying weird stars is not enough—we also need to interpret these signatures. Follow-up observations are certainly beneficial to such efforts but are frequently not available upon first identifying a peculiar signal. Pre-existing archival data (such as that taken by Gaia and 2MASS) as well as the shape of the light curve itself then become the immediate avenue for interrogating the signal in greater depth. The typical approach for interpreting light curve shape is to posit various hypotheses (e.g. comets, Dyson swarms, triangles) and compare their goodness-of-fit. But if none of these hypotheses are correct, then this ranking is ultimately of little value. A more agnostic approach is to use the shape of the light curve itself to generate a hypothesis.

This topic has received little attention in the prior literature but one recent study by Sandford & Kipping (2019) introduces the idea of inverting the 1D light curve morphology into a 2D reconstruction of the object's silhouette responsible for the transit—dubbed "shadow imaging" by the authors. This necessarily operates under the assumption that something did indeed transit the star, rather than say anomalous stellar activity. Further, as highlighted by the authors, the current algorithm developed is not exhaustive (a more optimal shape could be missed) and is a ripe problem for future software improvements.

Tabby's Star was an obvious target for this approach and a light curve inversion indicates the presence of semi-transparent material spanning a scale of several stellar radii with possible ring-like structures embedded (Sandford & Kipping 2019), thus independently confirming the previous work of Bodman et al. who came to the same conclusion via multiband photometry. This therefore demonstrates that light curve morphology is a ripe area for future study and could greatly aid our ability to classify anomalous signals in the future.